\documentclass[]{JHEP3}

\usepackage{epsfig}
\usepackage{amssymb}
\usepackage{amsmath}
\usepackage{mathrsfs}
 
\newcommand{\ensm}{\ensuremath}

\newcommand{\half}{\ensm{\frac{1}{2}}} \newcommand{\be}{\begin{equation}}

\newcommand{\gt}{\tilde{g}}
\newcommand{\ee}{\end{equation}} 

\newcommand{\dm}{\ensm{\partial_{\mu}}} \newcommand{\dn}{\ensm{\partial_{\nu}}}
\newcommand{\da}{\ensm{\partial_a}} \newcommand{\db}{\ensm{\partial_b}}
 
\newcommand{\asq}{\ensm{\alpha^2}}

\newcommand{\mn}{{\mu\nu}}
\newcommand{\nn}{\nonumber}
\newcommand{\B}{\begin{eqnarray}}
\newcommand{\E}{\end{eqnarray}}

\newcommand{\del}{\nabla}
\newcommand{\delm}{\nabla_\mu}
\newcommand{\deln}{\nabla_\nu}
\newcommand{\delt}{\tilde{\del}}

\newcommand{\psid}{\dot{\psi}}

\newcommand{\psidd}{\ddot{\psi}}

\newcommand{\kk}{\kappa_5^2}
\newcommand{\gab}{\overline{G}_{ab}}
\newcommand{\ab}{{ab}}
\newcommand{\vh}{\hat{V}}
\newcommand{\etad}{\dot{\eta}}
\newcommand{\etadd}{\ddot{\eta}}
\newcommand{\sg}{S_\textrm{g}}
\newcommand{\sbr}{S_\textrm{B}}
\newcommand{\sgh}{S_\text{GH}}
\newcommand{\sph}{S_\Phi}

\newcommand{\lag}{\mathcal{L}}

\title{Radion Dynamics in BPS Braneworlds}
\author{S.L. Webster\\Department of Applied Mathematics and Theoretical Physics, \\ Centre for Mathematical Sciences, \\ Wilberforce Road, \\ 
Cambridge, CB3 0WA, UK.\\{\tt S.L.Webster@damtp.cam.ac.uk}}
\author{A-C. Davis\\Department of Applied Mathematics and Theoretical Physics, \\ Centre for Mathematical Sciences, \\ Wilberforce Road, \\ 
Cambridge, CB3 0WA, UK.\\{\tt A.C.Davis@damtp.cam.ac.uk}}

\date{\today}

\abstract{We examine the moduli dynamics of a specific class of supergravity-inspired
BPS braneworlds, clarifying the role of bulk scalar fields in brane
collisions. The model contains as a special case the Randall-Sundrum model
both with and without a free, massless bulk scalar field.
Its low-energy effective theory is derived with a
moduli space approximation (MSA) and agrees with the corresponding results
derived elsewhere. Rather than stabilising the radion, we look at
cosmological evolution of the system stimulated by breaking the BPS
condition on the branes. We examine in detail the range of validity of the
MSA in both the RS and BPS case, paying particular attention to
the divergences that can arise during a collision of the branes. In
the absence of perturbations such
an event is finite in the RS model, and accurately described by the
low-energy effective theory. We demonstrate, however, that a collision
is divergent in the BPS case even with an exact FRW geometry.}
\preprint{DAMTP-2004-101}
\begin{document}

\section{Introduction}
Braneworld models for the Universe have recently been the subject of
considerable theoretical interest \cite{reviews}. Much of this research has been inspired by the work of Horava and
Witten \cite{hw}, who showed that the strong coupling limit of $E8\times E8$
heterotic String Theory could be described by an
eleven-dimensional supergravity with the eleventh dimension made
up of the orbifold $S^1/\mathbb{Z}_2$, i.e. an interval with reflection
symmetry about its endpoints. These endpoints define
ten-dimensional submanifolds, the branes, which sit at the
orbifold fixed points and define the boundaries of the spacetime.
The other six dimensions could consistently be compactified on a
Calabi-Yau threefold whose characteristic scale is considerably
smaller than the interbrane distance. This theory then led to many
toy models of spacetime as a five-dimensional manifold (the bulk) bounded by
two branes with a $\mathbb{Z}_2$ symmetry.

One of the most well-known of these is the Randall-Sundrum I model
\cite{rs}. Whereas the Horava-Witten model allows a large number
of fields to propagate in the bulk, this model has only a bulk
cosmological constant. The two boundary branes carry equal and
opposite tensions whose magnitudes can be fine tuned such that the
effective cosmological constants on the branes vanish, provided the
bulk cosmological constant is negative. The resulting low-energy theory can be derived in a
number of ways \cite{anne,kanno1,kanno2} and can be
formulated in terms of a single modulus field, the radion, representing the
proper distance between the branes. This four-dimensional low-energy
theory describes the physics as viewed by an observer confined to one
of the branes and, in the case of exact cosmological symmetry, turns
out to be perfectly well defined even when the size of the fifth
dimension vanishes, i.e. the branes collide.

The radion field is classically massless, representing the fact that
the brane positions are arbitrary when the brane tensions are fine tuned. This
is phenomenologically undesirable, since a massless scalar field is
not observed in nature. By detuning the brane tensions from their
preferred values one can generate a potential for the radion; one could imagine these detuning potentials on the
branes coming either from a more consistent quantum mechanical treatment of the scalar field
\cite{quantum} or from sort of SUSY-breaking mechanism. In this paper
we shall just insert it by hand.

We shall consider a supergravity-motivated generalisation \cite{anne2,anne} of the
Randall-Sundrum model where a scalar field $\Psi$ is allowed to propagate in
the bulk. Its potential $U$ replaces the Randall-Sundrum bulk
cosmological constant and the `superpotential' induced on the branes
$\vh$ replaces the tensions. Again the branes can be Minkowski if
$\vh$ and $U$ are related by a BPS condition, the analogue of the
Randall-Sundrum fine tuning. In fact, this model reduces exactly to
the Randall-Sundrum model if its free parameter $\alpha$, defined by
$\vh\varpropto\exp\alpha\Psi$, vanishes. This time, the four-dimensional
effective theory contains two moduli fields and, by detuning the brane
potentials, one can generate dynamics and cosmological solutions as before.

The cosmological consequences of this model have been explored in
detail in \cite{anne}. In this work we focus on brane collisions,
events which have been used recently in the literature to try to
illuminate and resolve the Big Bang singularity \cite{epk,bornagain}.
It is already known that perturbations diverge
during such an event; in this paper, we investigate how the simple
generalisation to BPS bulk scalar fields effects both the dynamics
of the radion and the regularity of the collision. Such bulk scalar
fields arise naturally in supersymmetric theories \cite{anne2}.

This paper is organised as follows. In \S\ref{a} we describe the model
and derive the exact projected Einstein equations on the brane. In the
cosmological context we obtain the Friedmann equation for the induced
FRW brane geometry. This is local (i.e. defined purely in terms of
quantities on the brane) apart from the well-known Weyl tensor
representing the gravitational effects of the bulk. Discussing the
low-energy effective theory in \S\ref{b} connects this quantity to
the moduli fields and allows one to write down an approximate, closed
set of equations for the motion of the branes. The dynamics in the
Randall-Sundrum case are reviewed in \S\ref{c}, with the
identification of the Weyl tensor explaining the finiteness of the
system through a brane collision. The more general case $\alpha\neq 0$
is studied in \S\ref{d}, showing that in all but the most contrived
case the scalar field will diverge during a brane collision.
The conclusions are summarised in \S\ref{e}.

\section{The Model}
\label{a}
A general action for the two-brane system is
\B
\label{5daction}
S&=&\frac{1}{2\kk}\int d^5x \sqrt{-g}\left[R-\frac{1}{2}\partial\Phi^2-U(\Phi)\right]\\&&+\int_1 d^4x\sqrt{-h^{(1)}}\left[2\frac{K}{\kk}-\hat{V}(\Phi)\right]
\nn\\&&+\int_2 d^4x\sqrt{-h^{(2)}}\left[2\frac{K}{\kk}+\hat{V}(\Phi)\right]\nn\E
ignoring for the time being the presence of possible matter actions
for the branes. $h^{(1)}_{ab}$ and $h^{(2)}_{ab}$ are the induced metrics on the positive-
and negative-tension branes respectively, $U$ and $\hat{V}>0$ are the
bulk and brane potentials and $K$ is the trace of the
extrinsic curvature $K_{ab}$ taking outward-pointing normals. The
factor of $\kappa_5^2$ ensures that $\Phi$ is dimensionless.

The index convention to be used is that
the metric signature is `mostly plus' $-++++$, $a,b,c...$ are five-dimensional
indices and $\mu,\nu,\rho...$ are four-dimensional, running from 0 to
3. The unusual factor of 2 in the Gibbons-Hawking sectors
is because of the $\mathbb{Z}_2$ orbifold nature of the bulk, and the five-dimensional integral in the action is taken to mean an integral
over \emph{two} copies of the bulk spacetime Region I between the two branes,
see Fig.\ref{bulkfig}. As described in the introduction the bulk
geometry is assumed to be $\mathbb{Z}_2$-symmetrical about the position of the two branes,
which simplifies the junction conditions. The $1$ and $2$ on the brane
integrals refer to positive
and negative tension respectively.

Variation of the action with respect to the metric and the field yield the usual
bulk equations and junction conditions for each
brane \cite{battye}. Considering the positive-tension brane (the analysis for the
other brane is equivalent with $\hat{V}\rightarrow-\hat{V}$), we
define $h_\ab\equiv h^{(1)}_\ab$ and find
\B
\label{geneinstein}
G_\ab&=&\half\da\Phi\db\Phi-\frac{1}{4}g_\ab\partial\Phi^2-\half
U(\Phi)g_\ab\\
\label{genkg}
\Box\Phi&=&\frac{dU}{d\Phi}\nn\\
\left[K_\ab\right]&=&-\kk \left(T_\ab-\frac{1}{3}Th_\ab\right)\nn\\
\left[n\cdot\partial\Phi\right]&=&2\kk\frac{d\vh}{d\Phi}\nn\E
where
\be
T_\ab=-\hat{V}(\Phi)h_\ab\ee
and $[X]$ means the jump in $X$ in the same direction as the
normal (hence both these quantities are invariant under $n\rightarrow -n$). Assuming $\mathbb{Z}_2$ symmetry the junction conditions become
\B
\label{genjump}
-K_\ab&=&-\frac{1}{6}\kk \hat{V}(\Phi)h_\ab\\
\label{fieldjump}
-n\cdot\partial\Phi&=&\kk\frac{d\vh}{d\Phi}\E
where quantities are evaluated at the edge of Region I, which the
normal points outward from as depicted in Fig. \ref{bulkfig}. The 4D Einstein tensor $\gab(h)$ is given by the standard result
\B
\label{G4}
\overline{G}_{ab}&=&\frac{2}{3}\left\{G_{cd}h^c_a h^d_b+\left(G_{cd}n^c
n^d-\frac{1}{4}G\right)h_{ab}\right\}\\&&+KK_{ab}-K_a^cK_{bc}-\half\left(K^2-K^{cd}K_{cd}\right)h_{ab}-E_{ab}\nn\E
where  $n^a$ is the spacelike
unit normal to the brane and $E_\ab$ is the (traceless) electric part of the Weyl tensor
\be
\label{eab}
E_\ab=C_{acbd}n^c n^d\nn\ee

\FIGURE{\includegraphics[width=9cm]{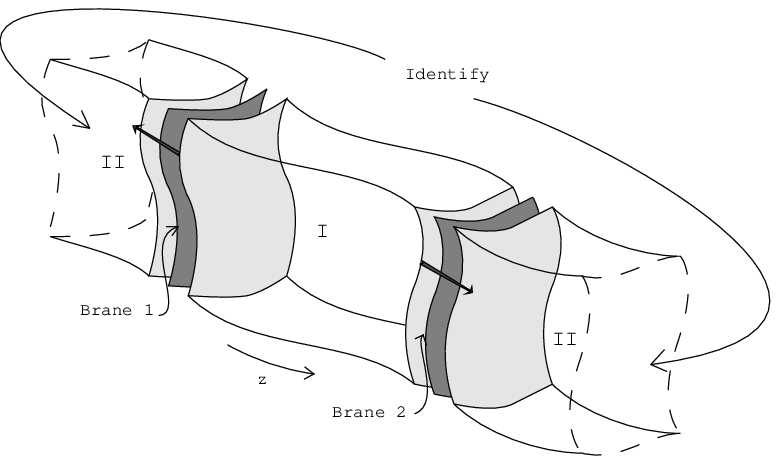}\caption{\textbf{Bulk and boundary structure}: The bulk topology is
  $\Sigma\times S_1/\mathbb{Z}_2$, with the branes sitting at the
  orbifold fixed points. There are therefore two identical copies of
  the bulk, Regions I \& II, with four boundary planes in all as shown,
  giving rise to two copies of the bulk action and of the GH boundary
  terms. There is only one copy of each brane worldvolume (and matter) action.}\label{bulkfig}}

From (\ref{geneinstein}),
\B
&&G_{cd}h^c_a h^d_b+\left(G_{cd}n^c
n^d-\frac{1}{4}G\right)h_{ab}\nn\\&&=\half\del_a\Phi\del_b\Phi-\frac{5}{16}\del\Phi^2
h_\ab+\frac{3}{16}\left(n\cdot\partial\Phi\right)^2h_\ab-\frac{3}{8}Uh_\ab\nn\E
where $\nabla$ is the covariant derivative of the induced metric $h$, and from (\ref{genjump}),
\be
\label{kk}
KK_{ab}-K_a^cK_{bc}-\half\left(K^2-K^{cd}K_{cd}\right)h_{ab}=-\frac{1}{12}\kappa_5^4
\vh^2h_\ab\nn\ee
Substituting all this back into (\ref{G4}) with the junction condition
(\ref{fieldjump}) we obtain the projected Einstein equation on the brane:
\B
\label{projeinstein}
\gab&=&\frac{1}{3}\del_a\Phi\del_b\Phi-\frac{5}{24}\del\Phi^2
h_\ab-E_\ab\\&&+\frac{1}{8}\kappa_5^4\left(\frac{d\vh}{d\Phi}\right)^2 h_\ab-\frac{1}{4}U h_\ab-\frac{1}{12}\kappa_5^4 \vh^2 h_\ab\nn\E
If matter were included on the brane via
\be
T_\ab\rightarrow -\vh(\Phi)h_\ab+\tau_\ab\ee
then there would also be the terms
\be
\label{quad}
...+\frac{\kappa_5^4}{6}\vh\tau_\ab+\kappa_5^4\pi_\ab\ee
where $\pi_\ab$ depends \emph{quadratically} on $\tau_\ab$. This has two
consequences; in order to recover the Friedmann equation we must live on the
positive-tension brane (at least at the classical level), and be at an
energy scale much less than the brane tension so that the resulting
$\tau_{00}^2$ term can be regarded as a small correction important
only in the early Universe. Note that the effective four-dimensional cosmological `constant' is
given by
\be
\Lambda_4=\frac{1}{4}U+\kappa_5^4\left[\frac{1}{12}\hat{V}^2-\frac{1}{8}\left(\frac{d\vh}{d\Phi}\right)^2\right]\nn\ee
which will vanish for potentials satisfying
\be
\label{susy}
U=\kappa_5^4\left[\frac{1}{2}\left(\frac{d\vh}{d\Phi}\right)^2-\frac{1}{3}\hat{V}^2\right]\ee
Such self-tuned potentials often arise in the context of supergravity
models where $U$ and $\vh$ would be derived from the same
superpotential \cite{anne2}. We shall for convenience refer to such potentials as
being `supersymmetric' and the function $\vh$ as the
superpotential. This relation between the bulk and brane potentials is a
generalisation of the usual Randall-Sundrum (henceforth
RS) fine tuning, which can be reproduced by taking $U=-2\Lambda$ and $\hat{V}=\sigma$; (\ref{susy}) then reduces to the familiar form
\be
\Lambda=-\frac{1}{6}\kappa_5^4\sigma^2\nn\ee

\section{Low-energy Effective Theory}
\label{b}
Whilst (\ref{projeinstein}) gives the exact Einstein equations for an
observer on the brane it has little predictive power, since it contains the
term $E_\ab$ which is not defined in terms of data on the brane. Such
explicit dependence on the bulk geometry is also present in the
Klein-Gordon equation (\ref{genkg}) which, when written out explicitly
in terms of four-dimensional covariant derivatives, will contain a
term of the form $\left(n\cdot\partial\right)^2\Phi$, not prescribed by
the value of the field on, and its derivatives along, the brane.

For this reason one seeks a four-dimensional effective theory in which
these non-local quantities are replaced by scalar `moduli fields'. Such low-energy effective theories can be derived in
many different, but essentially equivalent ways. Here we shall employ
the moduli space approximation; the result (\ref{action}) agrees with
that obtained by perturbative expansion \cite{gonzalo}.

Note that a four-dimensional description, low energy or otherwise,
in terms of an effective action cannot hope to reproduce a Friedmann
equation with the terms quadratic in the brane energy-momentum tensor
given by (\ref{quad}). For example \cite{KS2,paultba}, in the RS case, these can be
viewed as arising approximately from the trace anomaly in the CFT
defined on the branes in the context of the AdS-CFT correspondence,
and hence cannot be derived from the variation of an action. Higher
order derivative terms could be included in the effective action;
these will approximate then more and more closely the effects of
$E_\mn$ in the projected Einstein equations, but the quadratic
stress-energy terms cannot be obtained this way.

\subsection{BPS backgrounds}

The significance of supersymmetric (henceforth SUSY) potentials is
that they allow privileged configurations of the system
where the brane positions are arbitrary. These configurations are, in
a sense, a ground state of the system, with a high degree of
symmetry. We therefore look for solutions for the metric and the
scalar field which do not depend on the transverse directions
(i.e. are static and transversely homogeneous). We take Gaussian Normal coordinates away from the positive-tension brane
\be
ds^2=a(z)^2\eta_\mn dx^\mu dx^\nu +dz^2\nn\ee
where $z$ is the proper distance along the normal to the brane
(increasing towards the other brane) and $x^\mu$ parameterise the
flat, transverse foliations. Since we are assuming fine-tuning of the
brane tensions the transverse foliations are flat, hence the use of
$\eta_\mn$ above. Due to the symmetry of this `vacuum' configuration
the branes are at constant $z$. In the next section, when we relax the requirements of staticity and
homogeneity, we will consider more general coordinate systems.
Taking $\Phi(t,\mathbf{x},z)=\Phi(z)$, the Einstein and Klein-Gordon equations become:
\B
\frac{a'^2}{a^2}+\frac{a''}{a}&=&-\frac{1}{12}\Phi'^2-\frac{1}{6}U\nn\\
\label{bpssystem}
\frac{a'^2}{a^2}&=&\frac{1}{24}\Phi'^2-\frac{1}{12}U\\
\Phi''+4\frac{a'}{a}\Phi'&=&\frac{dU}{d\Phi}\nn\E
where $'=d/dz$. For potentials $U$ satisfying (\ref{susy}) this can be written as
\B
\label{eom1}
\left(\frac{a'}{a}\right)'&=&-\frac{1}{6}\Phi'^2\\
\label{eom2}\left(\frac{a'}{a}\right)^2&=&\frac{1}{24}\Phi'^2-\frac{1}{24}\kappa_5^4\left(\frac{d\vh}{d\Phi}\right)^2+\frac{1}{36}\kappa_5^4\vh^2\E
noting that there are only two independent equations in
(\ref{bpssystem}). An interesting family of solutions follow from the
first-order system
\be
\label{bps}
\frac{a'}{a}=-\frac{\kk\vh}{6}\ \ ,\ \ \Phi'=\kk\frac{d\vh}{d\Phi}\ee 
which can easily be seen to solve the full Einstein-Klein-Gordon
equations (\ref{eom1},\ref{eom2}). In the underlying supergravity theory
\cite{anne2} these are BPS configurations; the `no force' condition
between BPS branes manifests itself in that (\ref{bps}) \emph{implies} the junction conditions, which are given by
\be
\label{junction}
\left.\frac{a'}{a}\right|_1=-\left.\frac{\kk\vh}{6}\right|_1\ \ ,\ \ \Phi'|_1=\left.\kk\frac{d\vh}{d\Phi}\right|_1\ee
where the quantities are evaluated at the positive-tension
brane and we have used $K_\ab=g_\ab'/2$ for surfaces of constant $z$.
 In other words, there exist \emph{static} solutions of the system for arbitrary brane
positions. A solution of
(\ref{bps}) will from now on be referred to as a `BPS background'.
Note that the free parameters in these solutions are the value of the
scalefactor $a$ and the scalar field $\Phi$ on the positive-tension brane. 

Note that, in the metric junction condition (\ref{junction}), $a'/a$ is proportional to the extrinsic curvature on the
brane. Since the normal to the negative-tension brane must point in
the opposite direction for consistency (i.e. either both outward or
both inward), the corresponding junction condition there will pick up
a minus sign. Therefore, in order for the junction conditions to be
satisfied at both branes simultaneously, the negative-tension brane
must have tension
$\vh_-(\Phi)=-\vh_+(\Phi)$ as is assumed from the start in the action (\ref{5daction}).

As we shall see in the next section, it is desirable to detune the
brane potentials from their `supersymmetric' values. This will
generate a potential for the moduli fields in the four-dimensional
effective theory, which is desirable for both non-trivial dynamics and
some hope of their stabilisation. We shall insert this by hand,
\B 
\label{detune}
V_1(\Phi)&=& \vh(\Phi)+v(\Phi)\\
V_2(\Phi)&=&-\vh(\Phi)+w(\Phi)\nn\E
where $V_{1,2}$ are the tensions on the two branes, $\vh$ is the
supersymmetric value given by (\ref{susy}) and $v,w$ are small
perturbations, $v,w\ll\vh$. For definiteness, we shall consider only
exponential superpotentials,
\be
\label{superpotential}
\vh=\frac{6k}{\kk}e^{\alpha\Phi}\ee
which reduce to the RS model on taking the limit $\alpha\rightarrow 0$
whereupon the constant $k$, which has dimensions of inverse length, is the
curvature scale of the then AdS bulk.

\subsection{The Moduli Space Approximation}

In the previous subsection we derived a set of equations (\ref{bps})
whose solutions we will use as a ground state for our model. The
corresponding metric and scalar field profiles were static and
homogeneous, depending only on the bulk coordinate. This profile for
the scalar field and the metric satisfies the junction conditions
(\ref{junction}) at the branes provided the branes are parallel and
static. In order to examine small perturbations around this vacuum we allow
the brane positions to fluctuate as depicted in Fig. \ref{bulkfig},
and incorporate the graviton zero mode \cite{anne2} by generalising
the metric ansatz to
\be
\label{coords}
ds^2=dz^2+a(z)^2\gt_\mn(x) dx^\mu dx^\nu\ee
In this coordinate system the perturbed brane positions are given by
$z=z_1(x)$ and $z=z_2(x)$ (from now on, subscripts $1$ and $2$ will refer to evaluation at the
positive- and negative-tension branes respectively).
The radion, the proper distance between the
two branes, is given by $r(x)=z_2-z_1$. This is the local size of the
fifth-dimension; the only physical spacetime is that between the two branes.

We shall assume that any $x$-dependence is small, i.e. that transverse derivatives are
much smaller than normal derivatives. Also, we shall assume that the
SUSY-breaking potentials $v$ and $w$ (and matter actions were we to be
considering them) are small, to be consistent with the fact that we
must necessarily lose the quadratic terms in the Friedmann equation in
using an effective action, as discussed above. In this sense then the effective four-dimensional theory we will obtain is
only valid at low energies, specifically for
$v$,$w$,$\tau_\mn\ll\vh$. Radion fluctuations can be regarded as a small
perturbation around the BPS configuration; by substituting this
perturbed brane positions back into the action, keeping the BPS profile for the metric and
scalar field, we shall obtain an effective
theory governing the motion of the moduli fields $z_1$ and $z_2$.

For the superpotential (\ref{superpotential}) the BPS profile is given
by
\B
a(z)&=&\xi(z)^{1/6\asq}\nn\\
\label{bpsprofile}
\Phi(z)&=&-\frac{1}{\alpha}\log \xi(z)\\
\xi(z)&=&\left[6k\asq\left(z_0-z\right)\right]\nn\E
where $z_0$ is a constant of integration and we have assumed
$\alpha\neq 0$. As is generally the case
with self-tuned potentials there is a singularity in the bulk at
$z=z_0$. The theory will break down when this singularity lies in the
physical region of spacetime between the two branes \cite{anne}.

The action is given by
 \B
S_\textrm{full}&=&\sg+\sph+\sbr+\sgh\nn\\
\sg&=&\frac{1}{2\kk}\int d^5x\sqrt{-g}R(g)\\
\sph&=&\frac{1}{2\kk}\int d^5x\sqrt{-g}\left[-\frac{1}{2}\partial\Phi^2-U(\Phi)\right]\\
\label{s}
\sbr&=&\int_1 d^4x \sqrt{-h^{(1)}}\left(-\vh-v\right)\\&&+\int_2 d^4x \sqrt{-h^{(2)}}\left(\vh-w\right)\nn\\
\sgh&=&\frac{2}{\kk}\int_1 d^4x \sqrt{-h^{(1)}}K_1\ + \frac{2}{\kk}\int_2 d^4x\sqrt{-h^{{(1)}}} K_2\nn\\\E
remembering that
\be
\int d^5x\sqrt{-g}\ ... =2 \int d^4x \int_{z_1(x)}^{z_2(x)}dz \sqrt{-\gt}\ a^4\
...\nn\ee
Keeping the BPS profile for the metric and the field in place, we can
evaluate this term by term to obtain a four-dimensional effective
theory in terms of the $z_i$. For $v=w=0$ all terms not involving $x$-derivatives
will sum to zero since they would otherwise generate a potential term
for the moduli fields or a cosmological constant term, which both must
vanish in this fine-tuned case. Hence we can discard any terms not
involving transverse derivatives.

The result, expressed in terms of the induced metric on the
positive-tension brane $h^{(1)}_\mn\equiv h_\mn$, is derived in the
Appendix, and is given by

\be
S=\frac{1}{2k\kk}\int d^4x\sqrt{-h} \left[\Omega^2 R(h)-\gamma^{AB}\del\xi_A\cdot\del\xi_B-V\right]\nn\ee
with
\B
\Omega^2&=&\frac{1}{1+3\asq}\left[\xi_1-\xi_2\left(\frac{\xi_2}{\xi_1}\right)^{1/3\asq}\right]\nn\\
\gamma^{11}&=&\left[6\alpha^4\left(1+3\asq\right)
  \right]^{-1}\frac{1}{\xi_1}\left[3\asq+\left(\frac{\xi_2}{\xi_1}\right)^{1+1/3\asq}\right]\nn\\
\label{action}
\gamma^{12}&=&-\left[6\alpha^4 \right]^{-1}\frac{1}{\xi_1}\left(\frac{\xi_2}{\xi_1}\right)^{1/3\asq}\\\gamma^{22}&=&\left[6\alpha^4\right]^{-1}\frac{1}{\xi_2}\left(\frac{\xi_2}{\xi_1}\right)^{1/3\asq}\nn\\
V&=&2k\kk\left[v+w\left(\frac{\xi_2}{\xi_1}\right)^{2/3\asq}\right]\nn\E
in agreement with the result of \cite{gonzalo} (using different
normalisations for the fields). Here, and elsewhere, covariant
derivatives are taken with respect to the metric $h_\mn$ unless
otherwise specified. Finally we put the action into a more easily interpretable form, in
terms of the radion and value of the scalar field on the
positive-tension brane. For convenience we make the connection to the
radion via the approximate conformal factor $\omega^2$ relating the
two induced metrics,
\B
h^{(2)}_\mn&\approx&\omega^2 h^{(1)}_\mn\nn\\
\omega^2&=&\left(\frac{a_2(x)}{a_1(x)}\right)^2,\E
and define
\be
\psi(x)=1-\omega^{2(1+3\asq)}=1-\left(\frac{\xi_2}{\xi_1}\right)^{1+1/3\asq}\ee
so that the branes coincide for $\psi=0$
whilst $\psi=1$ signifies an infinite redshift between the two branes;
this could either mean their proper separation is infinite or that the
second brane has hit the singularity. The value of the scalar field on the positive-tension brane
is
\be
\eta(x)=-\frac{1}{\alpha}\log\xi_1\ee
giving our final action as
\B
S&=&\frac{1}{2\kappa^2}\int
  d^4x\sqrt{-h}e^{-\alpha\eta}\Big[\psi R(h)\nn\\&&\quad\qquad-\frac{3}{2\beta}\frac{1}{\left(1-\psi\right)}\del\psi^2-\frac{1}{2}\psi\del\eta^2\Big]\nn\\
&&\label{msa}-\int d^4x\sqrt{-h}\left\{v+\left(1-\psi\right)^{2/\beta}w\right\}\E
where
\be
\label{kappa}
\kappa^2=k\kk\beta\qquad\beta=1+3\asq\ee
Since the underlying action has five-dimensional coordinate
invariance, the choice of coordinate system used to perform the above
dimensional reduction cannot affect the final action, which is a
scalar quantity. Furthermore, if one changes the coordinate system
so that neither, one, or both of the branes are at fixed positions,
each component of the action is still separately invariant for the
same reason. This will, of course, depend on which four-dimensional
metric is being used, but if one is careful to express everything in terms of the same metric ($h_\mn$, for example) then it
is easy to check that each contribution to the total effective action
is unchanged. For example, one could use a coordinate system in which the branes are fixed, say at $y=0$ and $y=1$, via the transformation
$z=z_1(x)+r(x)y$. The scalar field modulus then enters as usual, being
defined as the value of the field at, for example, $y=0$, but this
time the radion enters directly in the metric,
\B
ds^2&=&r(x)^2 dy^2+2y\dm r(x) dx^\mu
dy\\&&+\left[\left(\frac{a(z(x,y))}{a_1(x)}\right)^2 h_\mn(x)+y^2\dm
r(x)\dn r(x)\right]dx^\mu dx^\nu\nn\E
The presence of these extra terms in the metric ensures that the
boundary terms still give precisely the same contribution.

\subsection{The RS Limit}
From (\ref{bpsprofile}), we see that
\be
\label{rslimit}
a(z)\sim a_0e^{-kz}\qquad\eta\equiv\Phi(z_1(x))\rightarrow 0\ee 
as $\alpha\rightarrow 0$, i.e. one recovers a AdS profile in the bulk
and the scalar field vanishes. Hence the RS model should be
understood as being recoverable at the level of the action by taking
$\alpha=\eta=0$:
\B
\label{rsaction}
S_{\textrm{RS}}&=&\frac{1}{2k\kk}\int d^4x\sqrt{-h}\Big[\psi
  R-\frac{3}{2\left(1-\psi\right)}\del\psi^2\nn\\
&&-2k\kk\left(v+\left(1-\psi\right)^2 w\right)\Big]\E

which is the standard action found in the literature \cite{kanno1}.
Note that setting $\eta=0$ is consistent with the equation of motion
obtained from the full action in the limit $\alpha\rightarrow 0$:
\be
\psi\Box\eta+\del\psi\cdot\del\eta=0\ee

If one takes $\alpha=0$ does not set $\eta$ to zero, the system
describes a RS braneworld with a free, massless bulk scalar field added by hand. With
a different normalisation for $\eta$, the resulting action is in
agreement with the result of \cite{kanno2}. 

\subsection{Scalar Degrees of Freedom as Goldstone Bosons}
The moduli fields have an interesting interpretation in terms
of symmetry breaking. In the RS case, the radion can be thought of as
the Goldstone boson associated with the breaking of translation
invariance. For example, if we fix the coordinate gauge invariance (in
the bulk direction) by defining the positive-tension brane to be at
$z=0$, there is still a continuous family of `vacuum states'
parameterised by the arbitrary position of the negative-tension brane
at $z=z_1$. Choosing a particular value of $z_1$ breaks this symmetry,
giving rise to a Goldstone mode which can be identified with the
radion. In the presence of the scalar field, one is also free to
choose the value of the scalar field e.g. at $z=0$, $\Phi_1$; there
are then two continuously deformable parameters $z_1$ and $\Phi_1$
parameterising the `vacuum manifold', giving rise to two massless
degrees of freedom when a specific choice of ground state is
made. When the tensions are detuned the moduli
develop a potential, and hence a mass, since they are no longer
Goldstone bosons - there is no longer a
continuous family of static solutions due to the violation of the junction conditions.

\section{Dynamics for RS}
\label{c}
In this section we re-derive the well-known result that tension
perturbations in RS generate a potential for the radion (see, for
example, \cite{bornagain,goldwise}) since it will be useful for the next Section where
we allow $\alpha\neq 0$. 
\subsection{Equations of motion}
The equations of motion which follow from the RS effective action for
the positive-tension brane (which we shall be working with for now on
unless otherwise specified) are
\B
\psi G_\mn&=&\delm\deln\psi-h_\mn\Box\psi\nn\\
&+&\frac{3}{2\left(1-\psi\right)}\left(\delm\psi\deln\psi-\half
h_\mn\del\psi^2\right)\nn\\
\label{rseomg}
&-&\kappa^2\ h_\mn\Big[v+\left(1-\psi\right)^2w\Big]\\
\Box\psi&=&-\half\frac{\del\psi^2}{1-\psi}-\frac{4\kappa^2}{3}\left(1-\psi\right)\Big(v+\left(1-\psi\right)w\Big)\nn\\
\label{rseompsi}
\E
where now $\psi$ is given by $\psi=1-\exp(-2kr)$. $\kappa$,as defined
from the action, is given by
\be
\label{rseffectivecoupling}
\kappa^2=k\kappa_5^2\ee
which is also the effective gravitational coupling on the brane; this 
can be identified directly from the projected Einstein equations in
the presence of matter (\ref{projeinstein},\ref{quad}), which give
\be
\label{kappa4}
\kappa_4^2=\frac{\kappa_5^4}{6}\vh\ee
For the RS case, $\vh=6k/\kk\ $, this is consistent with
(\ref{rseffectivecoupling}), i.e. $\kappa=\kappa_4$. In the absence of matter, the projected
Einstein equations for the positive-tension brane are
\be
\label{random1}
G_\mn=-E_\mn-\kappa^2v\ h_\mn\ee
allow the identification, to this level of
approximation, of
\B
E_\mn&=&h_\mn\frac{\Box\psi}{\psi}-\frac{1}{\psi}\delm\deln\psi\nn\\
&-&\frac{3}{2\left(1-\psi\right)}\frac{1}{\psi}\Big(\delm\psi\deln\psi-\half
h_\mn\del\psi^2\Big)\nn\\
\label{rsweyl}
&+&\kappa^2h_\mn\frac{1-\psi}{\psi}\Big(v+\left(1-\psi\right)w\Big)
\E
from (\ref{rseomg}). Note that the tracelessness of $E_\mn$ implies
the equation of motion (\ref{rseompsi}) for $\psi$. In fact \cite{KS2}, in the
absence of matter, the requirement that a single scalar field coupled to gravity yields Einstein
equations consistent with the tracelessness of the Weyl tensor
determines that, defining the scalar field to be the coefficient of
$R$ as above, the action \emph{must} be of the form
\B
S&\propto&\int d^4x\sqrt{-h}\Big[\psi
  R-\frac{3}{2\left(1-\psi\right)}\del\psi^2\nn\\
&+&A+B\left(1-\psi\right)^2\Big]\E
When the system is fine-tuned ($A=B=0$) one can write the
resulting Einstein equations as
\be
G_\mn=\kappa^2 T^{\psi}_\mn\nn\ee
and comparing with (\ref{random1}) one can identify the effective
energy-momentum tensor for $\psi$ with $-E_\mn$. The fact then that
$T^\psi_\mn$ must be traceless is sign of the underlying conformal
invariance of the RS effective action \cite{paultba}.

\subsection{Dynamics}

\FIGURE{\includegraphics[width=9cm]{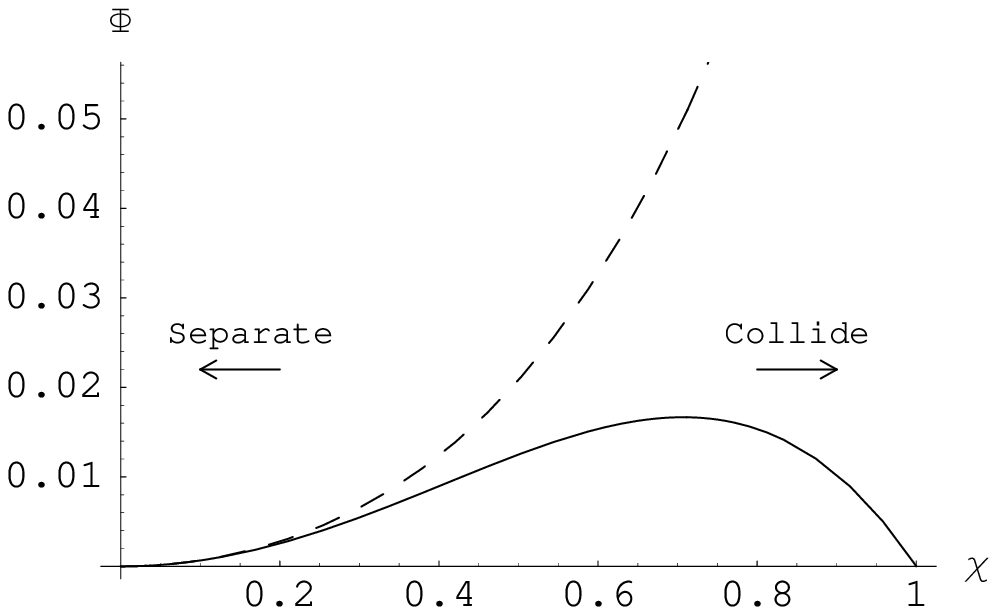}
\caption{\textbf{Effective potential for radion, RS case}: The dashed
  plot has $v=0.1$,$w=0.2$, and the potential drives the branes to
  infinite separation $\chi=0$. The solid plot has
  $v=0.1$,$w=-0.2$, giving an unstable fixed point at
  $\chi_c=1/\sqrt{2}$. The branes either collide or move apart
  depending on the initial conditions. $\kappa=1$ in both cases.}

\label{RSplot}}

The field $\psi$ has a non-standard kinetic term and, hence, its
dynamics are not
immediately obvious from (\ref{rseompsi}). Defining
\be
\chi=\sqrt{1-\psi}=e^{-kr}\ee 
we find
\be
\Box\chi=\frac{4\kappa^2}{3}\chi\left(v+\chi^2 w\right)\ee
i.e. the field $\chi$ moves in the
potential
\be
\label{RSpot}
\Phi(\chi)=\frac{4\kappa^2}{3}\Big(\frac{1}{2}\chi^2
v+\frac{1}{4}\chi^4 w\Big)\ee

This has a turning point at 
\be
\chi_c^2=-\frac{v}{w}\Leftrightarrow\psi_c=1+\frac{v}{w}\ee
which is only inside the physical range $0\leq\chi\leq 1$ for $v$ and
$w$ of opposite sign and $|v|\leq|w|$. This stationary point is
unstable for a de-Sitter positive-tension brane (i.e. positive
effective cosmological constant, $v>0$), i.e. an inflating brane, and stable
only when the brane is collapsing ($v<0$). Depending on the initial
conditions the branes will either be driven apart to $\chi=0\Leftrightarrow
r\rightarrow\infty$ or will collide, $\chi=1\Leftrightarrow
r=0$. However, if there is no stationary point in the potential
(for example, if $v>0$, when $v+w>0$) the branes will always be driven apart.
The potential (\ref{RSpot}) is plotted in Fig.\ref{RSplot} for two
different cases, and Figs.\ref{RSsim1} and \ref{RSsim2} show
explicitly the above behaviour.

\subsection{Brane Collisions}
As we have seen it is possible to produce an initial configuration
which will lead to a brane collision, which corresponds to
$\psi=0$. Whilst this appears in general to be a singular point of the
equations of motion, in the special case of an exact FRW induced
geometry the collision (in the Brane Frame, as used here) is
regular. We take a FRW metric with scalefactor
$a(t)$, curvature $K=0,\pm1$ and Hubble parameter $H=\dot{a}/a$.
The Bianchi Identity, 
\be\del^\mu G_\mn=0\Rightarrow\del^\mu E_\mn=0\Rightarrow E_{00}\propto a^{-4}\ee
gives, from (\ref{random1}),
\be
\label{rsfriedmann}
H^2+K/a^2=\frac{C}{a^4}+\frac{\kappa^2}{3}v\ee
for some constant dark radiation coefficient $C$.
Hence the Hubble
parameter does not diverge in the collision; the equations of motion
ensure that, despite appearing to be singular at $\psi=0$ from
(\ref{rsweyl}), $E_{00}$ actually only behaves as dark radiation. 
This is why we have chosen to work in the Brane Frame, rather than the
Einstein Frame defined by 
\B
\tilde{h}_\mn=\psi h_\mn\Rightarrow d\tilde{s}^2&=&\psi\left(dt^2-a^2
d\mathbf{x}^2\right)\nn\\
&\equiv&d\tilde{t}^2-\tilde{a}^2 d\mathbf{x}^2,\nn\E
which gives
\be
\tilde{H}\equiv\frac{1}{\tilde{a}}\frac{d\tilde{a}}{d\tilde{t}}=\frac{H}{\sqrt{\psi}}+\frac{\psid}{2\psi^{3/2}}\ee
which will, in general, diverge as $\psi\rightarrow 0$.

It is therefore possible to follow the evolution of the system through
a brane collision by using (\ref{random1}) (or, rather, its
derivative) in place of (\ref{rseomg}). Taking
$K=0$ for simplicity, the equations of motion for $\psi$ and $H$ are
\B
\label{rssimeqn1}
\dot{H}&=&-2H^2+\frac{2\kappa^2}{3}v\\
\label{rssimeqn2}
\psidd&=&-3H\psid-\half\frac{\psid^2}{1-\psi}\nn\\&+&\frac{4\kappa^2}{3}\left(1-\psi\right)\Big(v+\left(1-\psi\right)w\Big)\E
which are manifestly regular as $\psi\rightarrow 0$.
Note that, for $v>0$, (\ref{rssimeqn1}) has analytic solution
\be
H=\kappa\sqrt{\frac{v}{3}}\tanh\left(2\kappa\sqrt{\frac{v}{3}}t+\textrm{cnst}\right)\ee
i.e. the Universe rapidly approaches de Sitter space,
$H\rightarrow H_*=\kappa\sqrt{v/3}$. The evolution of the Hubble
parameter is completely independent of that of the radion and, in
particular, is not affected by $\psi=0$.

The initial value
of $H$ can be determined from the $00$-component of (\ref{rseomg}), which gives
\be
\label{rssimg00}
H^2+\frac{\psid}{\psi}H=\frac{\psid^2}{4\psi\left(1-\psi\right)}+\frac{\kappa^2}{3\psi}\Big(v+\left(1-\psi\right)^2w\Big)\ee
which in general has two solutions. Some possible evolutions of $H$
and $\psi$ for different initial conditions are given in
Figs.\ref{RSsim1} and \ref{RSsim2} which uses the potential of
Fig.\ref{RSplot} ($\psi_c=0.5, H_*/\kappa\approx0.183$). (\ref{rssimeqn1}) can also be obtained
as a linear combination of the three equations of motion obtained from
(\ref{rseomg}) and (\ref{rseompsi});  using these equations,
(\ref{rssimeqn2}), (\ref{rssimg00}) and 
\B
2\dot{H}^2+3H^2&=&H\frac{\psid}{\psi}-\frac{\psid^2}{4\psi(1-\psi)}\\&&-\frac{\kappa^2}{3\psi}\Big(\left(1-4\psi\right)v+\left(1-\psi\right)^2w\Big),\nn\E

gives the same results but more care
must be taken with the numerical integration to cope with the
divergence of some terms as $\psi\rightarrow 0$.

\FIGURE{\includegraphics[width=9cm]{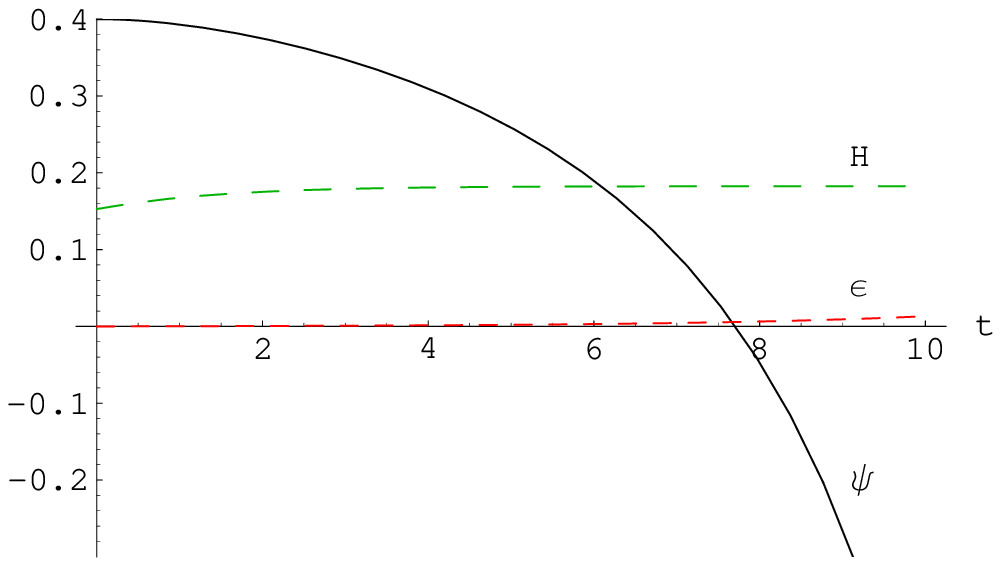}
\caption{\textbf{Brane collision in RS}: $\psi(t)$ and $H(t)$ are
  plotted for initial conditions leading to a brane collision
  ($\psi=0$). $H(t)\rightarrow H_*$ as expected.}
\label{RSsim1}}

\FIGURE{\includegraphics[width=9cm]{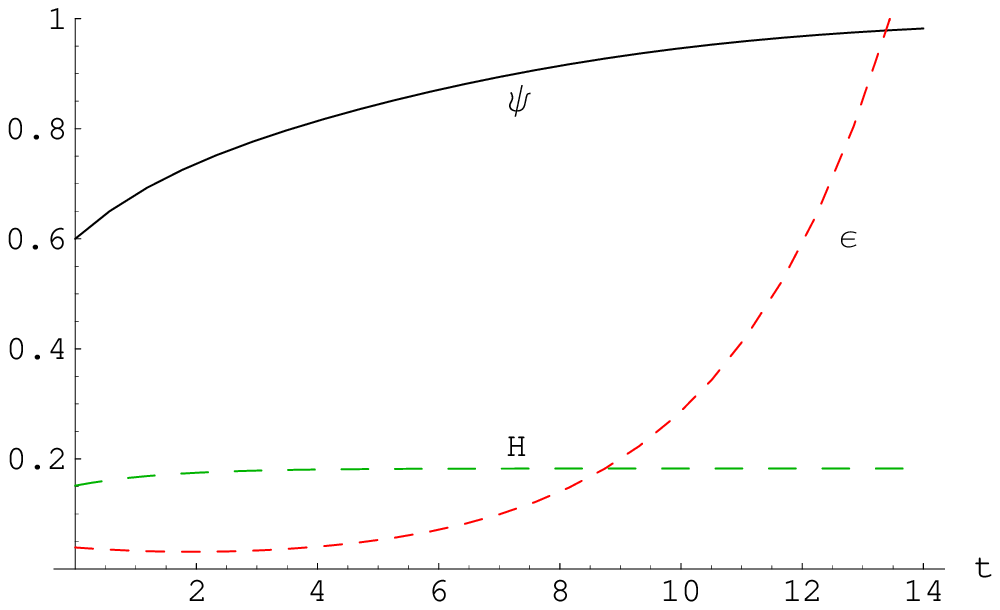}
\caption{\textbf{Branes driven apart}: $\psi(t)\rightarrow1$
  asymptotically for suitable choice of initial conditions. Note that
  $\epsilon$ ceases to be small as $\psi\rightarrow 1$}
\label{RSsim2}}

\subsection{Accuracy of the MSA}
As we have seen, the four-dimensional effective theory contains a
single modulus, $\psi$, related to the local radius of the orbifold dimension. 
The effective theory assumed that $\delt z_i^2/a_i^2\ll 1$, and we
need to check that these conditions are not violated during the evolution
of the system. It is actually sufficient to check that $\del
r^2/\omega^2\ll 1$. Taking a Gaussian Normal coordinate system where
the positive-tension brane lies at $y=0$, the metric is given by
\be
ds^2=dy^2+b(y)^2 h_\mn dx^\mu dx^\nu\ee
where the scalefactor $b(y)$ is normalised to $b(0)=1$, hence $h_\mn$
is indeed the induced metric. The same effective action
(\ref{rsaction}) is then obtained provided $\del r^2/\overline{b}^2\ll1$,
where $\overline{b}(x)$ is the scalefactor at the second brane,
$y=r(x)$. Clearly we can identify $\overline{b}(x)=\omega(x)$, giving
the above condition on $\del r^2$. In other words, although
only checking $\del r^2$ to be suitably small ignores the possibility
that $z_1$ and $z_2$ could oscillate wildly but coherently (i.e. that
$\del r^2$ small $\nRightarrow \del z_i^2$ small), all
that is of physical relevance is the separation, $r(x)$.
In terms of the modulus $\psi$, the condition is
\be\epsilon\equiv\left|\del
r^2/\omega^2\right|=\frac{1}{4k^2}\left|\frac{\del\psi^2}{\left(1-\psi\right)^3}\right|\ll
1\ee
As can be seen from Fig. \ref{RSsim2}, this suggests that the MSA will
become inaccurate as the branes move apart, $\psi\rightarrow 1$; a given
fluctuation will cease to be small as lengthscales shrink on the
negative-tension brane. As $\psi\rightarrow 0$ though, we just need
that $\del\psi^2\ll k^2$; the approximation appears to be valid
through the collision. In all the numerical simulations we shall
define our units by $k=\kappa_5^2=1$.

\section{Dynamics with Bulk Scalar}
\label{d}
Since the scalar field is assumed to take its BPS profile in the
background the effect of non-zero $\alpha$ on the dynamics is just to
alter some of the coefficients in the potential. In particular, we are
not attempting to produce a potential capable of stabilising the
radion as in $\cite{csaki,goldwise}$. The extra degree of freedom,
however, gives rise to key differences in the brane collision.

The gravitational coupling of matter on the brane is now given by
\be
\kappa_4^2(\eta)=k\kappa_5^2
e^{\alpha\eta}=\frac{\kappa^2}{\beta}e^{\alpha\eta}\ee
For simplicity we shall consider still the case where the tension perturbations $v$
and $w$ are constants; qualitatively identical results are obtained
for perturbations with the same functional form as the superpotential,
i.e. $v(\Phi)=\delta v\exp{\alpha\Phi}$.

\subsection{Equations of Motion}
The variation of (\ref{msa}) with respect to $h_\mn$, $\psi$ and
$\eta$ gives
\B
\psi G_\mn&=&\delm\deln\psi-h_\mn\Box\psi-\alpha\psi\left(\delm\deln\eta-h_\mn\Box\eta\right)\phantom{\Big)}\nn\\
&&-2\alpha\left(\del_{(\mu}\psi\del_{\nu)}\eta-h_\mn\del\psi\cdot\del\eta\right)\phantom{\Big)}\nn\\
&&+\alpha^2\psi\left(\delm\eta\deln\eta-h_\mn\del\eta^2\right)\nn\\
&&+\frac{3}{2\beta}\frac{1}{\left(1-\psi\right)}\left(\delm\psi\deln\psi-\half
h_\mn\del\psi^2\right)\nn\\
&&+\frac{\psi}{2}\left(\delm\eta\deln\eta-\half h_\mn\del\eta^2\right)\nn\\
&&-\kappa^2
\label{geneomg}
e^{\alpha\eta}h_\mn\left(v+\left(1-\psi\right)^{2/\beta}w\right)\\
R&=&-\frac{3}{2\beta}\frac{\del\psi^2}{\left(1-\psi\right)^2}+\half\del\eta^2+\frac{3\alpha}{\beta}\frac{\del\eta\cdot\del\psi}{\left(1-\psi\right)}\nn\\
&&-\frac{3}{\beta}\frac{\Box\psi}{\left(1-\psi\right)}-\frac{4\kappa^2}{\beta}e^{\alpha\eta}\left(1-\psi\right)^{2/\beta-1}w\nn\\
\psi R&=&\frac{3}{2\beta}\frac{\del\psi^2}{\left(1-\psi\right)}-\half\psi\del\eta^2+\frac{\del\psi\cdot\del\eta}{\alpha}+\frac{\psi}{\alpha}\Box\eta\nn\E
respectively ($\beta=1+3\asq$). Eliminating $R$ from the last two of these equations
yields
\B
\label{genboxpsi}
\Box\psi&=&\alpha\del\eta\cdot\del\psi-\frac{\del\psi^2}{2\left(1-\psi\right)}\nn\\
&&-\frac{4\kappa^2}{3}e^{\alpha\eta}\left[\left(1-\psi\right)v+\left(1-\psi\right)^{2/\beta}w\right]\\
\label{genboxeta}
\Box\eta&=&\alpha\del\eta^2-\frac{\del\eta\cdot\del\psi}{\psi}-\frac{3\alpha}{2\beta}\frac{\del\psi^2}{\psi\left(1-\psi\right)}\nn\\
&&+\frac{4\alpha}{\beta}\kappa^2 e^{\alpha\eta}v\E
consistent with (\ref{rseomg}) and (\ref{rseompsi}). The behaviour of
$\psi$ is qualitatively similar to the RS case, with the critical value of
$\psi$ modified to
\be
\psi_c=1-\left(-\frac{v}{w}\right)^\frac{1+3\asq}{1-3\asq}\ee
which is independent of $\eta$. Although there is no corresponding
stationary value of $\eta$ unless $\alpha=0$, (\ref{genboxpsi}) ensures
$\psi=\psi_c,\del\psi=0\Rightarrow\Box\psi=0$. Note that
(\ref{genboxpsi}) is regular as $\psi\rightarrow 0$.

The influence of the bulk scalar field is therefore not great on the effective
potential in which the radion moves; however, we shall see that it has
a more important role to play during a collision.

\FIGURE[p]{\includegraphics[width=9cm]{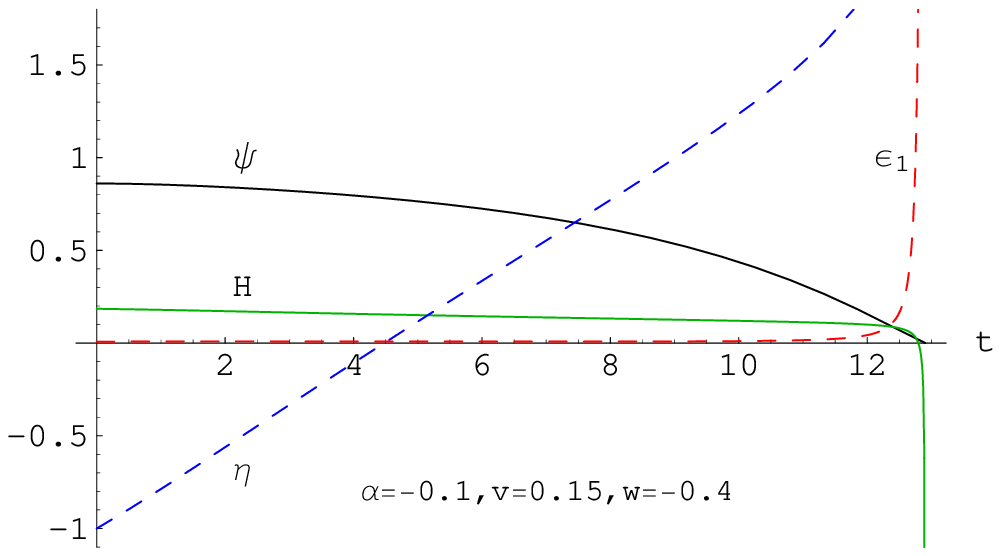}
\includegraphics[width=9cm]{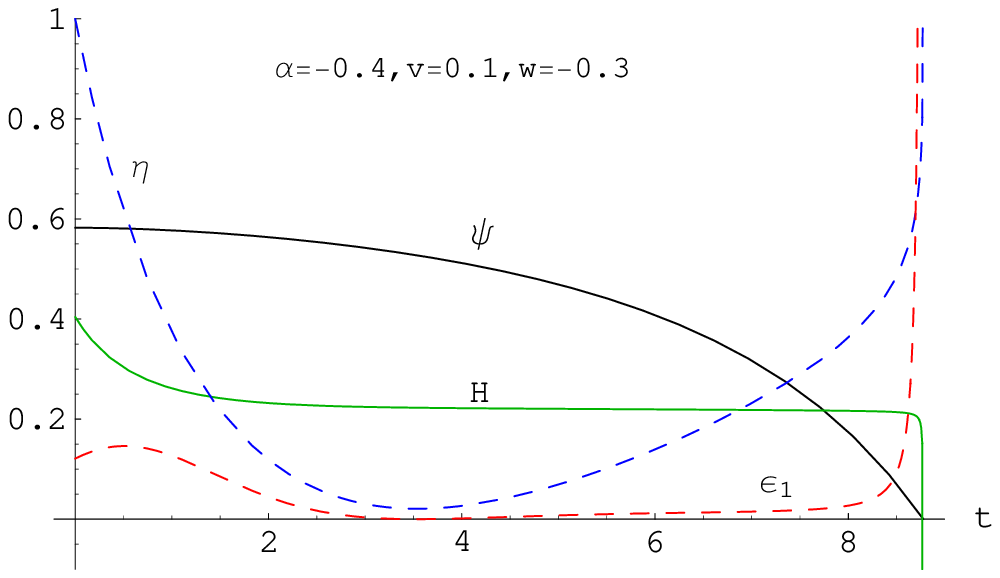}
\includegraphics[width=9cm]{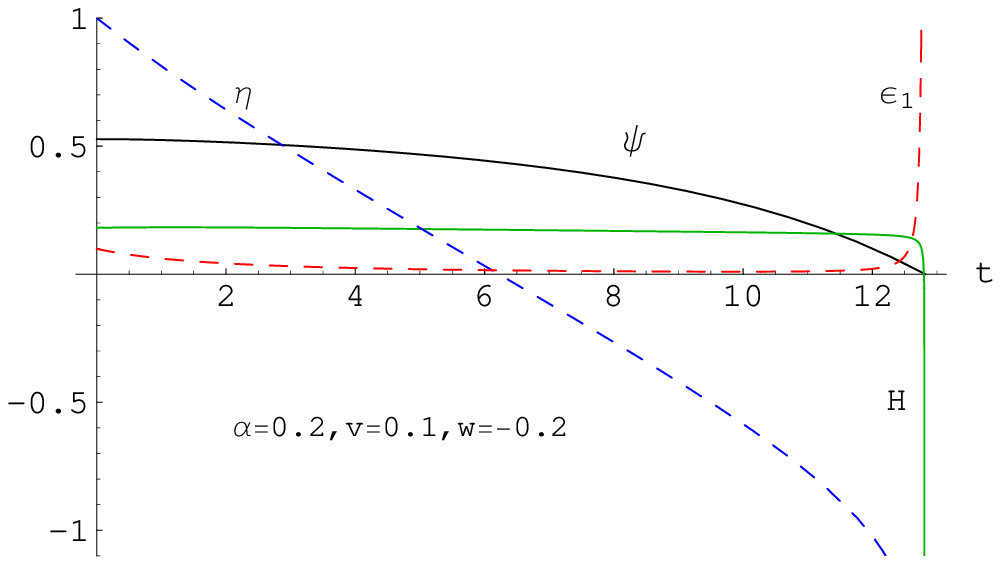}
\includegraphics[width=9cm]{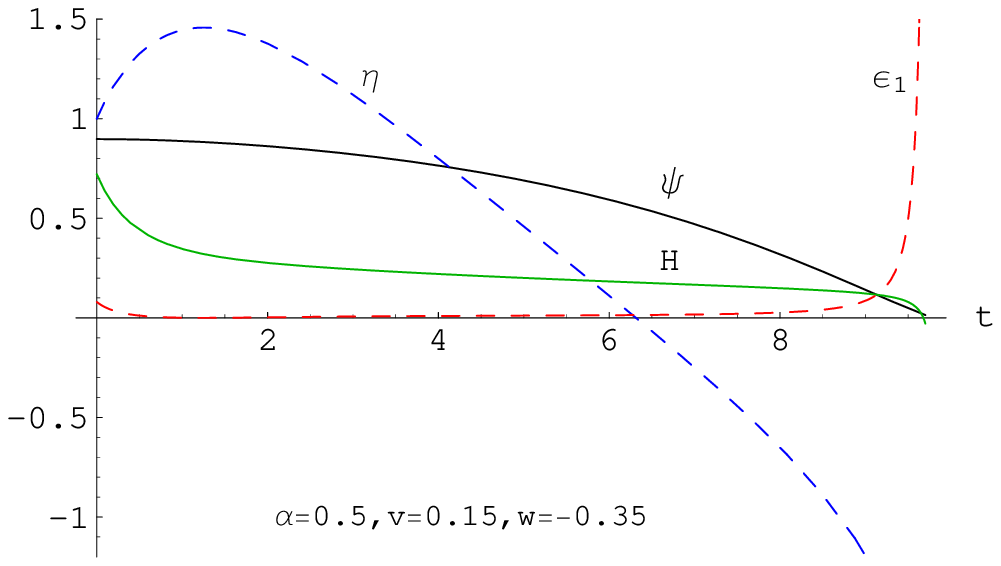}
\caption{\textbf{Divergence of $\eta$ as $\psi\rightarrow0$}: For four
different sets of parameters and initial conditions, $\eta$ can be
seen to diverge at the collision with opposite sign to $\alpha$. This
divergence causes a corresponding divergence in the parameter
$\epsilon_1$, signalling a breakdown of the effective theory description.}
\label{fullsim1}}

\subsection{Brane Collisions and Cosmological Evolution}
Taking again a flat FRW
metric, the Einstein equations (\ref{geneomg}) give the Hubble constraint 
\B
\label{geninith}
H^2&=&-H\frac{\psid}{\psi}+\alpha H\etad+\frac{1}{4\beta}\frac{\psid^2}{\psi\left(1-\psi\right)}+\frac{\etad^2}{12}\\
&&+\kappa^2 \frac{e^{\alpha\eta}}{3\psi}\Big(v+\left(1-\psi\right)^{2/\beta}w\Big)\nn\E
and the evolution equation
\B
2\dot{H}&=&-3H^2-\frac{\psidd}{\psi}-2H\frac{\psid}{\psi}+\alpha\etadd+2\alpha
H\etad\nn\\
&&-\Big(\asq+\frac{1}{4}\Big)\etad^2-\frac{3}{4\beta}\frac{\psid^2}{\psi\left(1-\psi\right)}+2\alpha\frac{\psid\etad}{\psi}\nn\\
&&+\kappa^2\frac{e^{\alpha\eta}}{\psi}\Big(v+\left(1-\psi\right)^{2/\beta}w\Big)\nn\\
&=&-3H^2+H\frac{\psid}{\psi}-\alpha H
\etad-\frac{\etad^2}{4}-\frac{1}{4\beta}\frac{\psid^2}{\psi\left(1-\psi\right)}\nn\\
\label{genoemg}
&&-\frac{\kappa^2}{3\psi}e^{\alpha\eta}\left[\left(1-\frac{4\psi}{\beta}\right)v+\left(1-\psi\right)^{2/\beta}w\right]
\E
using (\ref{genboxpsi}). Finally, substituting in from
(\ref{geninith}) gives the analogue of (\ref{rssimeqn1}):
\be
\label{finaleomh}
\dot{H}=-2H^2-\frac{\etad^2}{12}+\frac{2\kappa^2}{3\beta}e^{\alpha\eta}v\ee

\FIGURE{\includegraphics[width=9cm]{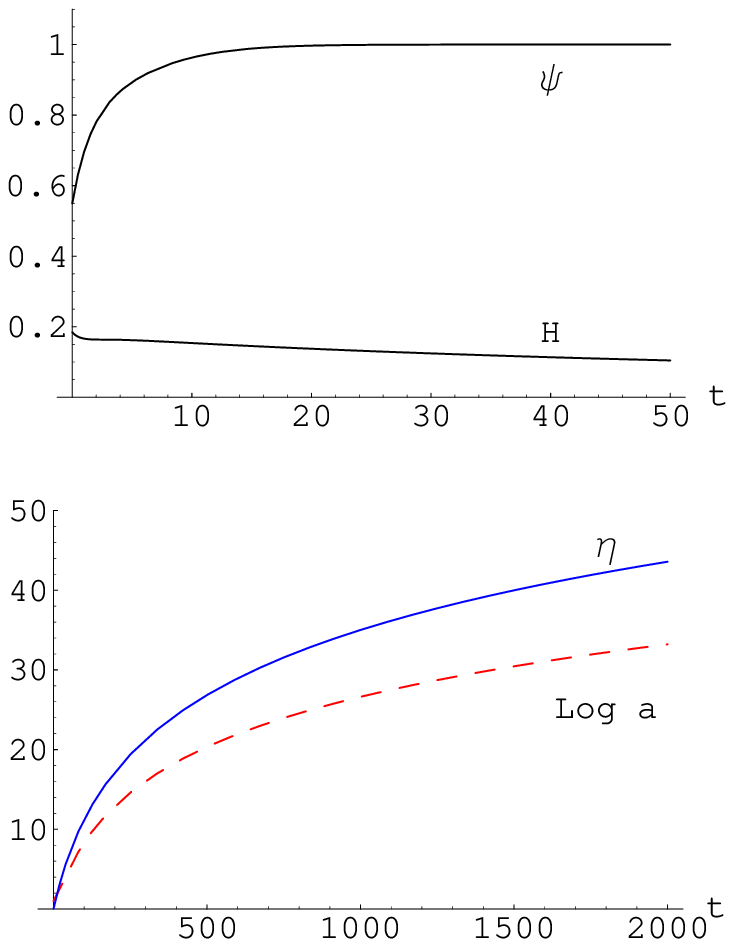}
\caption{\textbf{Late time cosmological evolution}:
  For suitable initial conditions $\psi$ is rapidly driven to unity
  and the branes move far apart. The scalar field grows
  logarithmically, the Hubble parameter is slowly driven to zero, and
  the Universe expands rapidly. Here $\alpha=-0.2$}  
\label{fullsim2}}

So again we have manifestly finite equations of motion for $H$ and $\psi$
as $\psi\rightarrow 0$. 
However, the equation of motion
(\ref{genboxeta}) is not free of divergences; if
$|\eta|\rightarrow\infty$ then, from
(\ref{finaleomh}), there will be a corresponding divergence
in $H$. Such behaviour is shown in Fig.\ref{fullsim1}.

For the $\alpha=0$ case, i.e. RS with a free, massless scalar field,
(\ref{genboxeta}) gives
\be
\delm\left(\psi\del^\mu
\eta\right)=0\quad\Rightarrow\quad\etad\propto\frac{1}{a^3\psi}\ee
Hence $\eta$ will tend to either a constant (if $\etad=0$, since the
field $\eta$ then has no time dependence and decouples) or to $\pm\infty$ as $\psi\rightarrow 0$ depending
on the initial sign of $\etad$.

In the general case it is difficult to make much progress
analytically. The numerical results for four sets of initial
conditions and values of $v,w$ and $\alpha$ are shown in
Fig.\ref{fullsim1}, appearing to show that $\eta$ diverges at the
collision with a sign opposite to that of $\alpha$ (a conclusion
supported by thorough numerical investigation, of which
Fig.\ref{fullsim1} is only a sample). This implies that
the tension on the branes, given by (\ref{superpotential}), vanishes
as the branes collide.

The conclusion is then that, although the scalar field
does not have much impact on the dynamics of the radion itself, it
will, in general, diverge during a collision (one could envisage a
situation where it would not, for example, by setting
$\alpha=\etad=0$, in which case $\eta$ would just remain constant and
the collision would be regular). This is similar to the evolution of
perturbations, which are found to diverge logarithmically in the RS
case \cite{bornagain}, a feature common to scalar-tensor theories when the gravitational constant changes sign.

When the initial conditions are such that the branes do not collide,
$\psi$ is rapidly driven to 1. $\eta$ is then
approximately governed (for $\alpha\neq 0$) by
\be
\etadd+3H\etad\sim\alpha\etad^2-\frac{4\alpha}{\beta}\kappa^2
e^{\alpha\eta}v\ee
The Hubble constant is slowly driven to zero whilst
$|\eta|$ grows approximately logarithmically (a full numerical solution is given in Fig.\ref{fullsim2}
for $\alpha=-0.2$). There is no analytic
solution to the system even if one approximates $\psi=1$; one can, however, predict the late time behaviour
\be
\label{eta}
\eta-\eta_0\sim-\frac{2}{\alpha}\log\left|1+t-t_0\right|\ee
with the timescale for the transition to the logarithmic behaviour
increasing without limit as $\alpha\rightarrow 0$. This is clearly demonstrated in Fig. \ref{etasim}.

\FIGURE{\includegraphics[width=9cm]{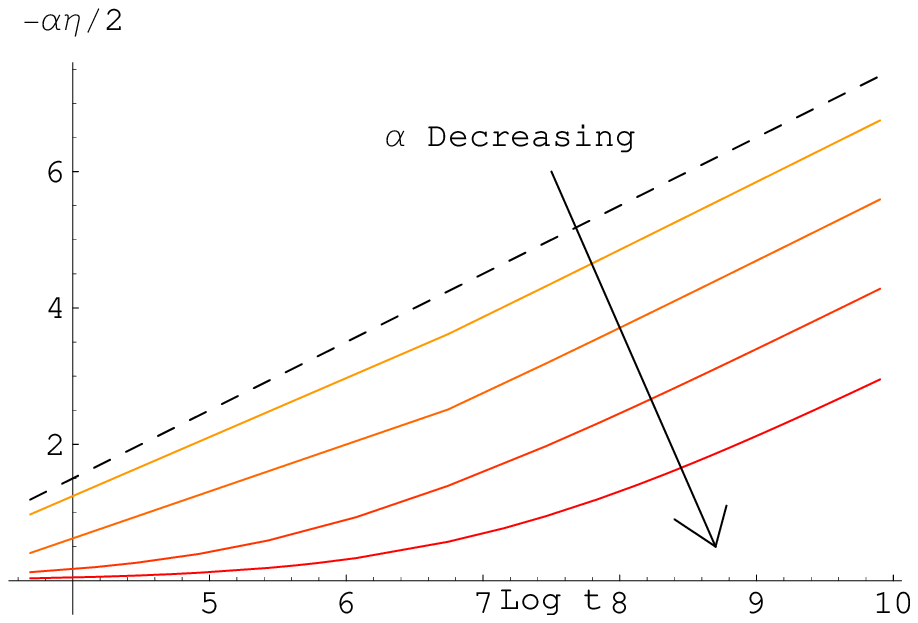}
\caption{\textbf{Late-time $\eta(t)$}:
The evolution of $-\alpha\eta/2$ as $\psi\rightarrow 1$ is
plotted, with $\alpha=0.4,-0.2,0.1,-0.05$ from top to bottom. The
dashed fiducial line with unit gradient demonstrates the late-time
behaviour $\eta\sim -\frac{2}{\alpha}\log t$. As $\alpha$ decreases it
takes longer for the system to settle into this behaviour.} 
\label{etasim}}

\subsection{A Caveat}
As before, the self-consistency of the effective theory can be checked by
monitoring the size of the quantities
\B
\epsilon_1\equiv\left|\frac{\del z_1^2}{a_1^2}\right|&=&\frac{e^{\left(1/{3\alpha}-2\alpha\right)\eta} }{36 k^2\asq}\left|\del\eta^2\right|\nn\\
\epsilon_2\equiv\left|\frac{\del
  z_2^2}{a_2^2}\right|&=&\frac{e^{\left(1/{3\alpha}-2\alpha\right)\eta} }{36 k^2\asq}\left(1-\psi\right)^{-3/(1+3\asq)}\nn\\&&\qquad\qquad \times\big|\left(1-\psi\right)\del\eta+3\alpha\del\psi\big|^2\nn\\&\sim&\epsilon_1\qquad\mathrm{as}\ \psi\rightarrow 0\nn
\E
However, due to the divergence of $\eta$, these are not necessarily
small during the collision, as can be seen in Fig.\ref{fullsim1}. The
reason for this divergence, and the reason for its absence in the RS
case, can be understood as follows. In our coordinate system in which both
the bulk and the scalar field are static, one can
interpret a divergence in $\eta$ as a divergence, or a tending to the
singularity $z_*$ depending on the sign, of the coordinate $z$
of the positive-tension brane. In the RS case, the two coordinates
diverge in the collision. However, the effective theory can be
expressed solely in terms of
the difference between the two values, the radion, which remains
finite. In the more general case, the
divergence of the second modulus cannot be removed in such a way. The
breakdown of the effective theory can also be understood by rederiving the effective action in a Gaussian-Normal
coordinate system about the positive-tension brane, analogously to the
procedure used in \cite{kanno2}. One performs a perturbative expansion
about a static ground state \cite{gonzalo}, requiring transverse derivatives of the scalar field
to be small. However, this procedure
will break down when this condition is no longer met, in
particular when $\eta$ diverges during the collision.

Whilst then we cannot trust the MSA right at the moment of collision,
we have a firm numerical tool for identifying dynamically the region
in which it is, indeed, valid. We can use the MSA to identify the
approach to a collision and, although we have not ruled out the
possibility of higher-order effects repelling the branes again, it
seems likely that a collision will then take place, accompanied by a
divergence in the bulk scalar field. 

\subsection{Einstein vs. Brane Frame}
In this paper we have chosen to work in the Brane Frame, defined by
the induced metric on the positive-tension brane, since this has the
best regularity properties at the brane collision. For late-time
evolution the Einstein Frame is most often used in the literature;
from the action (\ref{msa}), this can be read off as
\be
g_\mn=e^{-\alpha\eta}\psi h_\mn\ee
As $\psi\rightarrow 0$ the two metrics will give wildly different
physics, but in the RS case, with or without the scalar field, the two metrics become identical as
$\psi\rightarrow 1$, so the distinction between the two frames becomes
unimportant. However, for $\alpha\neq 0$, we see from 
(\ref{eta}) that
\be
e^{-\alpha\eta}\psi\propto \frac{1}{t^2}\quad\mathrm{as}\quad
\psi\rightarrow 1
\ee
so that, even for late times, there is still a distinction between the
two frames. The cosmological evolution of the moduli fields in the
Einstein Frame are investigated in detail in \cite{anne}.
 
\section{Conclusions}
\label{e}
In this paper, we studied the evolution of the moduli fields of a
particular class of SUGRA-inspired braneworld models, containing the
RS model as a special case, with a scalar field in the bulk. In
contrast to previous work we worked in the frame of the induced metric
of one of the branes in order to study collisions, which are always
singular in the Einstein Frame.
After discussing the significance of `supersymmetric' pairings of brane interaction and
bulk scalar field potentials from a fully five-dimensional viewpoint,
we derived an effective theory to discuss the low-energy motions of
the system about a BPS ground state configuration with small
SUSY-breaking potentials.
It was found that, for this specific class of braneworlds, the presence of a bulk scalar field has little
qualitative effect on the dynamics of the radion itself, i.e. the
motion of the branes through the bulk. However, even in the absence of any
perturbations around an FRW induced geometry, its contribution to the
brane Friedmann equation causes the Hubble parameter to diverge at the
collision even in the Brane Frame.
By monitoring the size of certain functions of the field variables
which are required to be small in the derivation of the effective
action, we can identify the regions in which the MSA itself is
valid. Whilst collisions are well-described by it in the RS case, the
presence of the bulk scalar field can causes it to break down just
before the collision.
Although the RS model can be used to construct toy models whose
regularity at the collision gives a link between pre- and post-Big
Bang phases, where the Big Bang is identified with the moment of
collision, this work implies that any attempt to generalise the model
to include bulk scalar fields encounters singularities.
Recent developments on modelling brane collision in the context of
M-Theory \cite{malcolm} suggest that, with more care, these
divergences might turn out to be removable; purely within the context
of general relativity, however, they appear to be unavoidable.

\begin{center}
\textbf{Acknowledgements}
\end{center}

This work is supported in part by PPARC. The authors would like to thank
Malcolm Perry, Carsten van de Bruck, Gonzalo Palma and Paul McFadden
for many helpful discussions. SLW would also like to thank Claudia de
Rham and Philippe Brax for comments on an earlier version of this paper.

\section{Appendix}

Firstly, the Ricci scalar decomposes as
\be
R(g)=-8\frac{a''}{a}-12\frac{a'^2}{a^2}+\frac{\tilde{R}}{a^2}\nn\ee
where $\tilde{R}$ is the Ricci scalar of the metric
$\gt_\mn$. Writing $S=\int d^4x \sqrt{-\gt}\ \lag$, we find
\B
\lag_\text{g}&\supset&\frac{\tilde{R}}{\kk}\int_{z_1(x)}^{z_2(x)}dz
\ a^2(z)\nn\\
\label{lagg}
&=&\frac{\xi_1^{1+1/3\asq}-\xi_2^{1+1/3\asq}}{2k\kk\left(1+3\asq\right)}\tilde{R}
\E
where $\xi_i=\xi(z=z_i(x))$. The scalar field Lagrangian contains no
$x-$derivatives (as shown in \cite{wavefn} there is no additional
perturbation around the BPS field configuration that needs to be taken
into account) and hence makes no overall contribution.

The moduli kinetic terms come from the boundary action. For the
$i$-brane the induced line element
and outward-pointing normal are given by
\B
ds_i^2&=&\left(a_i^2\gt_\mn+\dm z_i\dn z_i\right)dx^\mu dx^\nu\\&&\equiv
h^{(i)}_\mn dx^\mu dx^\nu\nn\\
n^{(i)}_a dx^a&=&(-1)^i\frac{1}{\sqrt{1+\delt z_i^2/a_i^2}}\left(dz-\dm z_i dx^\mu\right)\nn\E
\vspace{-5mm}
giving
\be
\label{deth}
\sqrt{-h^{(i)}}\approx a_i^4\sqrt{-\gt}\left(1+\frac{1}{2 a_i^2}\delt z_i^2\right)\ee
where $\delt$ is the covariant derivative of $\gt_\mn$. Here, and from
now on, we are assuming that $\del z_i^2\ll a_i^2$, i.e. that the brane
fluctuation lengthscale is much larger than that of the bulk
curvature. Then, from (\ref{s}),
\be
\label{lagb}
\lag_B\supset -a_1^4\ v-a_2^4\ w+\frac{3k}{\kk}\left(\frac{1}{\xi_2}a_2^2\delt
z_2^2-\frac{1}{\xi_1}a_1^2\delt z_1^2\right)\ee
where we have from (\ref{superpotential}) and (\ref{bpsprofile}) that
\be
\vh(\Phi)=\frac{6k}{\kk}\frac{1}{\xi}\nn\ee
and we have assumed that the SUSY-breaking potentials $v$ and $w$ are also
small. Finally, we need to compute the Gibbons-Hawking boundary term involving
the extrinsic curvatures of the two branes. Taking outward-pointing
normals again,
\B
K_i&=&\left.\frac{1}{\sqrt{-g}}\da\left(\sqrt{-g}\ n_{(i)}^a\right)\right|_{z=z_i(x)}\nn\\
&\approx&(-1)^i\left.\frac{1}{a^4}\left(a^4\left(1-\half\frac{\delt
  z_i^2}{a^2} \right)\right)'\right|_{z=z_i(x)}\nn\\&&-(-1)^i\frac{1}{\sqrt{-\gt}
  a_i^4}\dm\left.\left(a^2\sqrt{-\gt}\ \gt^\mn\dn
z_i\right)\right|_{z=z_i(x)}\nn\\
&=&-(-1)^i\frac{4k}{\xi_i}\nn\\&&+\frac{(-1)^i}{a_i^2}\left[k\frac{\delt
  z_i^2}{\xi_i}-\frac{1}{a_i^2}\left.\left\{\delt_\mu\left(a^2\delt^\mu z_i\right)\right\}\right|_{z=z_i(x)}\right]\nn\E
From (\ref{deth}) and (\ref{s}) it then follows that
\B
\lag_\mathrm{GH}&\supset&\frac{2}{\kk}\left[a_1^2\frac{k}{\xi_1}\delt
  z_1^2+\left.\left\{\delt_\mu\left(a^2\delt^\mu
  z_1\right)\right\}\right|_{z_1(x)}\hspace{-3mm}-\left(\textrm{1}\leftrightarrow\textrm{2}\right)\right]\nn\\
&=&\frac{2}{\kk}\left[a_1^2\frac{k}{\xi_1}\delt
  z_1^2-2a_1\delt^\mu z_1\delt_\mu a_1-\ \left(\textrm{1}\leftrightarrow\textrm{2}\right)\right]\nn\\
\label{laggh}
&=&\frac{6k}{\kk}\left(a_1^2\frac{\delt
  z_1^2}{\xi_1}-a_2^2\frac{\delt z_2^2}{\xi_2}\right)\E
where integration by parts has been used in the second line, and using
  (\ref{bpsprofile}). It is straightforward to show that the omitted terms above some to
  zero as expected, leaving an effective action
\B
S&=&\frac{1}{2k\kk}\int d^4 x \sqrt{-\gt}\ \tilde{\Omega}^2
  \tilde{R}+\frac{1}{6\alpha^4}\left[\xi_1^{1/3\asq-1}\delt\xi_1^2\right.\nn\\&-&\left.\xi_2^{1/3\asq-1}\delt\xi_2^2\right]-\tilde{V}\\
\tilde{\Omega}^2&=&\frac{1}{1+3\asq}\left[\xi_1^{1+1/3\asq}-\xi_2^{1+1/3\asq}\right]\nn\\
\tilde{V}&=&\xi_1^{2/3\asq}\ v+\xi_2^{2/3\asq}\ w\nn\E
As expected, the potential for the moduli fields vanishes for $v=w=0$;
when supersymmetry is unbroken (i.e. (\ref{susy}) holds), the moduli
fields are massless. Next we write the action in terms of the induced metric on the
  positive-tension brane, $h^{(1)}_\mn\equiv h_\mn$. To this order it is sufficient to
  use the approximation
\be
h_\mn=a_1^2\gt_\mn+...\ee
Using
\be
\tilde{R}=a_1^2\left(R(h)-6a_1\Box a_1^{-1}+...\right)\ee
we finally obtain, after integration by parts, the action as given by (\ref{action}).

\end{document}